\documentclass[11pt]{revtex4-1}

\usepackage{graphicx}

\begin{document}

\title{Dynamical System Analysis for a phantom model}
\author{Nilanjana Mahata\footnote {nilanjana\_mahata@yahoo.com}}
\author{Subenoy Chakraborty\footnote {schakraborty.math@gmail.com}}

\affiliation{Department of Mathematics, Jadavpur University, Kolkata-700032, West Bengal, India.}



\begin{abstract}
The paper deals with a dynamical system analysis related to  phantom cosmological model . Here gravity is coupled to phantom scalar field having scalar coupling function and a potential. The field equations are reduced to an autonomous dynamical system by a suitable redefinition of the basic variables and assuming some suitable form of the potential function. Finally, critical points are evaluated, their nature have been analyzed and corresponding cosmological scenario has been discussed.
\end{abstract}

\maketitle



\section{Introduction}
The standard cosmology seems to overcome the blow due to recent observational evidences  [1,2] by introducing an exotic matter [3] having large negative pressure - the dark energy (DE). Cosmological constant [4-8] is the natural choice as DE but it is not well accepted due to its extreme fine tuning and coincidence problem [9]. As a result, it is natural to  consider dynamical DE having (negative) variable equations of state [10]. People have gone one step further to consider matter with the equation of state parameter $w <-1 $ which is not ruled out by observational data and moreover necessary to describe the current acceleration of the universe. Such a model is known as phantom [ 11] DE model whose various aspects have been studied in [12-15].\\
However, despite of a lot of works (for reviews see the references 3,7, 8 and 16) still we are completely in the dark about the source of the DE, it is only characterized by its negative pressure. Although the $\Lambda$CDM model suggests the accelaration as a final stage but there is no priori reason why the acceleration of the universe will be the final stage. Other DE models may have different fate of the universe. In particular, in phantom DE model the energy density may increase with time while Hubble parameter and the curvature diverges in finite time and the final fate of the universe is  a big-rip singularity [12, 13, 17]. In the present paper, we consider a phantom model of the universe, where gravity is coupled to a scalar field with a scalar coupling function and a potential. The basic equations are presented in section II and also the evolution equations are converted into an autonomous system by suitable transformation of the basic variables. Section III  and IV deals  with the analysis of the critical points of the system  for 3D and 2D respectively while stability criteria for the  system has been studied in section V. Cosmological implications of the equilibrium points has been discussed in section VI. Finally, there are short discussion and concluding remarks in section VII.
\section{ Basic Equations}
The equation of a scalrar field coupled to gravity with coupling parameter, a function of the scalar field $\phi$ has the form
\begin{equation}
 A = \int d^{4}x\sqrt{-g}~ [~\frac{R}{2k^{2}}-\frac{1}{2}\lambda(\phi)g^{\mu\nu}(\nabla_{\mu}\phi)(\nabla_{\nu}\phi)-V(\phi)]+ S_{m}
\end{equation}
where $k^{2} = 8\pi G$ is the gravitational coupling, $ \lambda(\phi)$, the coupling parameter is chosen as an arbitrary function of $\phi$, $V(\phi)$ is the potential of the scalar field and $ S_{m}$ is the action of the matter field which is chosen as the cold dark matter (DM) in the form of dust. For flat FRW model, if we vary the action with respect to the metric then we have the Einstein equations:
\begin{equation}
 H^{2} = { \frac{k^{2}}{3}}(\rho_{m} + \rho_{\phi}),  ~~~~     and~~~~
   \dot{H}  = - \frac{k^2}{2}(\rho_{m} + \rho_{\phi} + p_{\phi}),
\end{equation}
where
\begin{equation}
\rho_{\phi} = \frac{1}{2}\lambda(\phi)\dot{\phi}^{2}+ V(\phi)~~~~ and~~~~ p_{\phi} = \frac{1}{2}\lambda(\phi)\dot{\phi}^{2} - V(\phi)
\end{equation}
are respectively the energy density and thermodynamic pressure of the scalar field. It is assumed that the scalar field does not interact with cold DM and the energy conservation relations take the form
\begin{equation}
\dot{\rho_{m}}+ 3H\rho_{m} = 0
\end{equation}
\begin{equation}
\dot{\rho_{\phi}}+ 3H(\rho_{\phi} + p_{\phi}) = 0
\end{equation}
i.e the scalar field satisfies the evolution equation
\begin{equation}
\lambda(\phi)\ddot{\phi} + \frac{1}{2}\lambda'(\phi)\dot{\phi}^{2}+ 3H\lambda(\phi)\dot{\phi}+\frac{\partial V}{\partial\phi} = 0
\end{equation}
As the evolution equations (i.e equations (2) and (6)) are non-linear and complicated in form so we shall discuss the cosmological evolution through qualitative analysis i.e we shall transform the evolution equations into an autonomous dynamical system to perform the phase space and stability analysis by introducing the auxiliary variables x and y [3, 18-21].\\
In general, the cosmological equations of motion reduce to a self-autonomous system :

$\overrightarrow{X}' = \overrightarrow{f}(\overrightarrow{X})$ where the column vector $ \overrightarrow{X}$ is constituted by the auxiliary variables  and $ \overrightarrow{f}(\overrightarrow{X})$ is the corresponding column vector  of the autonomous system and prime denotes differentiation with respect to logarithm of scale factor. The critical points $ \overrightarrow{X_{c}}$ are obtained from $ \overrightarrow{X}' = 0$ i.e $\overrightarrow{f}(\overrightarrow{X_{c}}) = 0 .$ In order to study the
 stability criteria of the equilibrium (i.e critical) points, we use the 1st order perturbation technique and obtain the matrix equation $ \overrightarrow{U}' = M\overrightarrow{U}$
where the column vector $\overrightarrow{U}$ denotes the perturbation of the variables and matrix M contains the coefficients of the perturbation equations. Then the eigen values of M characterize the nature of the critical point and the stability is determined by the conditions :  $ Tr M < 0  $  and  $ det M > 0 .$\\
In the present problem the auxiliary variables are defined as [19]

\begin{equation}
x = \frac{\sqrt{\lambda(\phi)}}{\sqrt{6}H}\dot{\phi} ~~~~~, ~~~~~ y = \frac{\sqrt{V(\phi)}}{\sqrt{3}H} ~~~ and ~~~~z = \frac{\sqrt{6}}{\phi}
\end{equation}

Then the first Friedman equation in (2) shows the interrelation between the new variables  x and y as
\begin{equation}
x^{2}+ y^{2} +\Omega_{m} = 1
\end{equation}
where $\Omega_{m}= \frac{\rho_{m}}{3H^{2}}$ is the usual density parameter for cold dark matter. Further, using these new variables the above evolution equations can be written as \\
 $~~~~~~~~~~~~~~~~~~~~~~~~~~~~~~~~~~~~~~~~~~~~~~~~~~ x'= \frac{3}{2} x (x^{2}- y^{2}-1)- \alpha_{1}(z) y^{2}$
 \begin{equation}~~~~~~
 y'= y [ \alpha_{1}(z) x +\frac{3}{2}(x^{2}- y^{2}+ 1)]
\end{equation}
$~~~~~~~~~~~~~~~~~~~~~~~~~~~~~~~~~~~~~and ~~~~~~z' = -\frac{xz^{2}}{\sqrt{\lambda_{1}(z)}}$\\
where $  \alpha_{1}(z) = \alpha(\phi) = \sqrt{\frac{3}{2}}\frac{V'(\phi)}{V(\phi)\sqrt{\lambda(\phi)}}~~~ and ~~~~ \lambda_{1}(z) = \lambda(\phi)$.\\
 Note that the above first order system of non-linear differential equations (9) can be considered as a 3D autonomous system. In the following section we shall study the autonomous system with some specific choice of the potential $ V(\phi)$(exponential) and the coupling function $\lambda(\phi) $ (exponential and power law). \\
The effective equation of state for the scalar field (i.e $ p_{\phi}= (\nu_{\phi}- 1)\rho_{\phi})$ is given by
\begin{equation}
   \emph{} \nu_{\phi} = \frac{2x^{2}}{x^{2}+ y^{2}}
\end{equation}
and the net effective equation of state ( $\omega_{eff}$) has the expression
\begin{equation}
   \omega_{eff} = \frac{p_{\phi}}{\rho_{\phi} + \rho_{m}} = x^{2}- y^{2}
\end{equation}
For cosmic acceleration $\omega_{eff} < -\frac{1}{3}$ is required.
The density parameter for the scalar field has the expression
\begin{equation}
   \Omega_{\phi} = \frac{\rho_{\phi}}{3H^{2}}= x^{2}+ y^{2}
\end{equation}
It should be noted that the physical region in the phase plane is constrained by the requirement that the energy density be non-negative i.e $\Omega_{m}\geq 0$. So equation (8) restricts the dependent variables x and y to be on the circular cylinder $ x^{2}+ y^{2}\leq 1$. Further, geometrically equation (8) represents a paraboloid in $ ( \Omega_{m}, x ,y)$-state space and it is possible to divide the 3D-state space into the following invariant sets:\\
  A : $ \Omega_{m} > 0 ~~~~ and~~~~ ~~~~~~ x^{2}+ y^{2} < 1, $ non-vacuum 2D \\
 $ ~~B :  \Omega_{m} = 0 ~~~~ and ~~~~~~~~~~~x^{2}+ y^{2} = 1, $ no fluid matter 1D
\section{ Analysis of Critical points : 3D autonomous system }
\subsection{ Exponential potential and exponential coupling function}
 Here we choose
\begin{equation}
 V(\phi) = V_{0}e^{\mu(\phi)}~~~~ and ~~~~\lambda(\phi)= \lambda_{0}e^{\nu\phi}
 \end{equation}
 Then the autonomous system (9) becomes\\\\\\
 $~~~~~~~~ ~~~~~~~~~~~~~~~~~~~~~~~~~~~~~~~~~~~~~~~x'= \frac{3}{2} x (x^{2}- y^{2}-1)- \alpha_{0} y^{2}e^{-\frac{b}{z}}$  
\begin{equation}
 y'= y [ \alpha_{0}e^{-\frac{b}{z}} x +\frac{3}{2}(x^{2}- y^{2}+ 1)]
\end{equation}
$~~~~~~~~~~~~~~~~~~~~~~~~~~~~~~~~~~~~~~~~and ~~~z' = -\frac{xz^{2}e^{-\frac{b}{z}}}{\sqrt{\lambda_{0}}}$\\
with $~~\alpha_{0} = \sqrt{\frac{3}{2\lambda_{0}}}\mu~~$ and $~~ b = \sqrt{\frac{3}{2}}\nu~~$.\\
The critical points of the present autonomous system  are $ C_{1}(0,0,0)$, $C_{2}(0,1,0),C_{3}(0,-1,0)$, $ C_{4}(1,0,0), \\
C_{5}(-1,0,0)$ and $C_{6}(0,0,z)$.\\

The 3x3 matrix for linearized perturbation  is given by
 \[
   A =
  \left[ {\begin{array}{ccc}
   \frac{9}{2}x_{c}^2-\frac{3}{2}y_{c}^2 - \frac{3}{2}    & ~~~~~~~~~~~   -3y_{c}x_{c}- 2\alpha_{0}y_{c}e^{-\frac{b}{z_{c}}} & ~~~ \frac{b}{z_{c_{}}^{2}}\alpha_{0}y_{c}^{2}e^{-\frac{b}{z_{c}}} \\
   3y_{c}x_{c} + \alpha_{0}y_{c}e^{-\frac{b}{z_{c}}} & ~~~~~~~\alpha_{0}x_{c}e^{-\frac{b}{z_{c}}} + \frac{3}{2}((x_{c}^2 + 1)-\frac{9}{2}y_{c}^2  &  -\frac{b}{z_{c}^{2}}\alpha_{0}x_{c}y_{c}e^{-\frac{b}{z_{c}}}\\
    -\frac{z_{c}^{2}}{\sqrt{\lambda_{0}}}e^{-\frac{b}{z_{c}}} & 0 & -\frac{x}{\sqrt{\lambda_{0}}}e^{-\frac{b}{z_{c}}}(b + 2z_{c})\\
  \end{array} } \right]
\]
where ($ x_{c},y_{c},z_{c}$) is an equilibrium (critical) point of the  above autonomous system. As at least one of the eigen values for each critical point is zero so all are non-hyperbolic in nature. Hence we can not investigate the local stability of the system , only we can find the dimension of the stable manifold ( if exists) by applying the centre manifold theorem [22,23]. Table I shows the critical points and information regarding them.\\
From table I we find that the points $C_{1}$ and $C_{6}$ indicate that universe is completely dominated by dark matter and late time acceleration is not possible . They correspond to flat non accelerating universe completely dominated by dark matter having only one stable manifold.
$ C_{2}$  have stable manifold of 2D and represents  De-Sitter accelerating universe completely dominated by dark energy.  $ C_{3}$ denotes contracting universe ( $ y < 0 $ ) so physically not meaningful. $ C_{4}$ and $C_{5}$ does not have stable manifold. Both are dominated by dark energy. None of the above  points match with the observations and can not solve coincidence problem.

\subsection{ Exponential potential and power law coupling parameter}

For the choice $ V(\phi) = V_{0}e^{\mu(\phi)}~~~~ and ~~~~\lambda(\phi)= \lambda_{0}\phi^{n}$, ($ V_{0}, \lambda_{0}$ and n as constant ), the autonomous system (9) becomes\\

 $~~~~~~~~~~~~~~~~~~~~~~~~~~~~~~~~~~~~~~ ~~~x'= \frac{3}{2} x (x^{2}- y^{2}-1)- \alpha_{2} y^{2}z^{\frac{n}{2}}$
\begin{equation}
 y'= y [ \alpha_{2}z^{\frac{n}{2}} x +\frac{3}{2}(x^{2}- y^{2}+ 1)]
\end{equation}
$~~~~~~~~~~~~~~~~~~~~~~~~~~~~~~~~~~~~~~~~and ~~~z' = - \lambda_{2}xz^{2 + \frac{n}{2}}$\\
with $ \alpha_{2} = \alpha_{0}6^{-\frac{n}{4}}$ and $ \lambda_{2} = 6^{-\frac{n}{4}} \frac{1}{\sqrt{\lambda_{0}}}$.\\
Here also $ C_{i}, i = 1,2,... 6$ are the critical points of the autonomous system (15).\\
 The  linearized perturbation matrix for the above autonomous system takes the form
 \[
   A =
  \left[ {\begin{array}{ccc}
   \frac{9}{2}x_{c}^2-\frac{3}{2}y_{c}^2-\frac{3}{2} & ~~~~~~~~~~~~~~~~-3y_{c}x_{c}- 2\alpha_{2}y_{c}z_{c}^{\frac{n}{2}} & ~~~ -\frac{n}{2}\alpha_{2}y_{c}^{2}z_{c}^{\frac{n}{2}-1} \\
   3y_{c}x_{c} + \alpha_{2}y_{c}z_{c}^{\frac{n}{2}} & ~~~~~~~\alpha_{2}x_{c}z_{c}^{\frac{n}{2}} + \frac{3}{2}(x_{c}^2 + 1)-\frac{9}{2}y_{c}^2  &  \frac{n}{2}\alpha_{2}x_{c}y_{c}z_{c}^{\frac{n}{2}-1}\\
    -\lambda_{2}z_{c}^{\frac{n}{2}+2} & 0 & -(\frac{n}{2}+2)\lambda_{2}x_{c}z_{c}^{\frac{n}{2}+1}\\
  \end{array} } \right]
\]

 \begin{table}
 \caption{\emph{Equilibrium points, related  parameters and Eigen values } }
 \begin{tabular}{|c|c|c|c|c|c|c| p{1 in}|p{1 in}|}

 \hline
   Equilibrium point & x & y & z & $  \Omega_{m} $ & $ \Omega_{\phi}$ & $ \omega_{eff}$ & Stable manifold & Eigen Values \\\hline
 $ C_{1}$ & 0 & 0 & 0 & 1 & 0 & 0 & $ 1 D $ & $ 0, \frac{3}{2},-\frac{3}{2} $ \\\hline
 $ C_{2}$ & 0 & 1 & 0 & 0 & 1 & -1 &  $2 D $ & $ 0, -3, -3 $ \\\hline
 $ C_{3}$ & 0 & -1 & 0 & 0 & 1 & -1 & $2 D $ &  $ 0, -3, -3 $ \\\hline
 $ C_{4}$ & 1 & 0 & 0 & 0 & 1 & 1 & no stable manifold & $ 0, 3, 3 $ \\\hline
 $ C_{5}$ & -1 & 0 & 0 & 0 & 1& 1 & no stable manifold & $ 0, 3, 3 $  \\\hline
 $ C_{6}$ & 0 & 0 & z & 1 & 0 & 0  & 1 D & $ 0, -\frac{3}{2},\frac{3}{2} $ \\\hline

 \end{tabular}
 \end{table}

 As before the critical points are non-hyperbolic in nature . Thus all the critical points are same as case A (given by table I) and the implications are also same as case A.
\section{ Analysis of Critical points : 2D autonomous system }
If the potential function $ V(\phi)$ and the coupling function $\lambda(\phi)$ are chosen such that $\alpha(\phi)$ is either a constant or a function of x or y (or both) then we have 2D autonomous system as \\
 ~~~~~$~~~~~~~~~~~~~~~~~~~~~~~~~~~~~~~~~~ ~~~~~~~~~~~~~~x'= \frac{3}{2} x (x^{2}- y^{2}-1)- \alpha(\phi) y^{2}$
  \begin{equation}
 y'= y [ \alpha(\phi) x +\frac{3}{2}(x^{2}- y^{2}+ 1)]
\end{equation}
Auxiliary variable z is not considered here.
We shall now analyze this autonomous system for some choices of $ V(\phi)$ and $\lambda(\phi)$ so that\\
 i) $\alpha $ is a constant , ii) $ \alpha = \frac{1}{2x}$ and iii)  $ \alpha = - \frac{1}{2x}$
\subsection{ $\alpha $ is a constant }
 In this case the potential function of the scalar field and the  coupling function $\lambda(\phi)$ are related by the relation:
 \begin{equation}
 V(\phi) = V_{0}~~ exp~ [~\sqrt{\frac{2}{3}}~\alpha \int\sqrt{\lambda(\phi)}d\phi~]
 \end{equation}
 with $ V_{0} $ , a constant of integration.
 This inter relation between $ V(\phi)$ and $\lambda (\phi)$  is possible for the following realistic choices of the potential function\\
 a)$V(\phi) = V_{0} e^{\mu\phi}, \lambda(\phi)= constant $\\
  b) $ V(\phi) = V_{0} exp[\sqrt{\frac{2\lambda_{0}}{3}}\frac{\alpha}{l+1}\phi^{l+1}], \lambda(\phi)= \lambda_{0}\phi^{2l}, (l\neq -1)$\\
  c) $ V(\phi) = V_{0} \phi^{\sqrt{\frac{2\lambda_{0}}{3}}\alpha}, \lambda(\phi)=\frac{ \lambda_{0}}{\phi^{2}}. $\\ Thus both exponential and power law form of potential is possible for this choice of $\alpha$. \\
  In this case the above autonomous system (16) has seven critical points which are presented in table II and are discussed below:
 \begin{table}
 \caption{\emph {Equilibrium points with their nature}}
 \begin{tabular}{|c|c|c|c|c|c|p{2 in}|}
 \hline
 Equilibrium point & x & y & $  \Omega_{m} $ & $ \Omega_{\phi}$ & $ \nu_{\phi}$ & Nature \\\hline
 $ E_{1}$ & 0 & 0 & 1 & 0 & undefined & saddle point \\\hline
 $ E_{2}$ & 1 & 0 & 0 & 1 & 2 & saddle if $ \alpha < - 3$,  unstable node if $ \alpha > - 3$\\\hline                              $ E_{3}$ & -1 & 0 &0 & 1 & 2  & saddle if $ ~\alpha~ > ~3$,  unstable node if $ ~~\alpha ~< ~ 3$\\\hline

 $ E_{4}$ & $ -\frac{\alpha}{3}$& $\sqrt{1-\frac{\alpha^2}{9}}$& 0 & 1 & $\frac{2\alpha^2}{9}$ & stable node if $ \alpha^2 < \frac{9}{2}$, saddle point if $\frac{9}{2} < \alpha^2 < 9 $  \\\hline
 $ E_{5}$ & $- \frac{\alpha}{3}$& $ - \sqrt{1-\frac{\alpha^2}{9}}$ & 0 & 1 &$\frac{2\alpha^2}{9}$ & stable node if $ \alpha^2 < \frac{9}{2}$, saddle point if $\frac{9}{2} < \alpha^2 < 9 $  \\\hline
 $ E_{6}$ & $- \frac{3}{2 \alpha}$ & $ \frac{3}{2\alpha} $ & $ 1 -\frac{9}{2\alpha^2} $ & $ \frac{9}{2\alpha^2} $ & 1 & saddle point if $\alpha < \pm\frac{3}{\sqrt{2}}$, stable node if $\frac{9}{2} < \alpha^2 \leq \frac{36}{7} $, stable focus if $ \alpha^2 > \frac{36}{7} $\\\hline
 $ E_{7}$ & $- \frac{3}{2\alpha}$& $ -\frac{3}{2\alpha} $ & $ 1 -\frac{9}{2\alpha^2} $& $\frac{9}{2\alpha^2}$ &1 & saddle point if $\alpha < \pm\frac{3}{\sqrt{2}}$, stable node if $\frac{9}{2} < \alpha^2 \leq \frac{36}{7} $, stable focus if $ \alpha^2 > \frac{36}{7} $ \\\hline
 \end{tabular}
 \end{table}

 \emph{Equilibrium point $ E_{1}$ : x = 0, y = 0}\\
 The physical parameters namely $ \Omega_{m}, \Omega_{\phi},$ and $ \nu_{\phi}$ are presented in table II. As
  $ \Omega_{\phi}= 0 $ ( i.e $ \Omega_{m} = 1)$ so the universe is completely dominated by dark matter. Note that in this case $ \nu_{\phi}$ is completely unspecified and the results hold independently of its value.

\emph{Equilibrium point $ E_{2}$ : x = 1, y = 0  and $E_{3}$ : ~$ x = -1, y = 0 $ }\\
 $ E_{2}$ corresponds to scalar field dominated universe with values of the cosmological parameter in table II. The scalar field behaves as relativistic stiff fluid. The equilibrium point $E_{3}$ represents the same epoch with identical values of the cosmological parameters.

\emph{Equilibrium point $ E_{4}$ : x = $ -\frac{\alpha}{3}$, y = $\sqrt{1-\frac{\alpha^2}{9}}$ and $ E_{5}$ : x = $ -\frac{\alpha}{3}$, y = $ - \sqrt{1-\frac{\alpha^2}{9}}$}\\
To have these critical points to be realistic, $ \alpha $ is restricted to be $ -3 < \alpha < 3 $. The values of the cosmological parameters shown in table II, indicate that they  have identical behaviour and both correspond to scalar field dominated universe (DM is absent). Further, the scalar field behaves as a perfect fluid and depending on the various choices of $\alpha $ , critical points correspond to universe from radiation era to $\Lambda$-CDM model.

\emph{Equilibrium point $ E_{6}$ : x = $ -\frac{3}{2\alpha}$, y = $\frac{3}{2\alpha}$ and $ E_{7}$ : x = $ -\frac{3}{2\alpha} $, y = $- \frac{3}{2\alpha}$}\\
Here phantom scalar field behaves as dust. So essentially, both the points represent universe dominated by dark matter.\\
The critical points ( given in table II ) and direction field near those critical points are given in Fig 1 and Fig 2 for two different values of $\alpha $.

\begin{figure}
\begin{minipage}{.45\textwidth}

 \includegraphics[width = 1.0\linewidth]{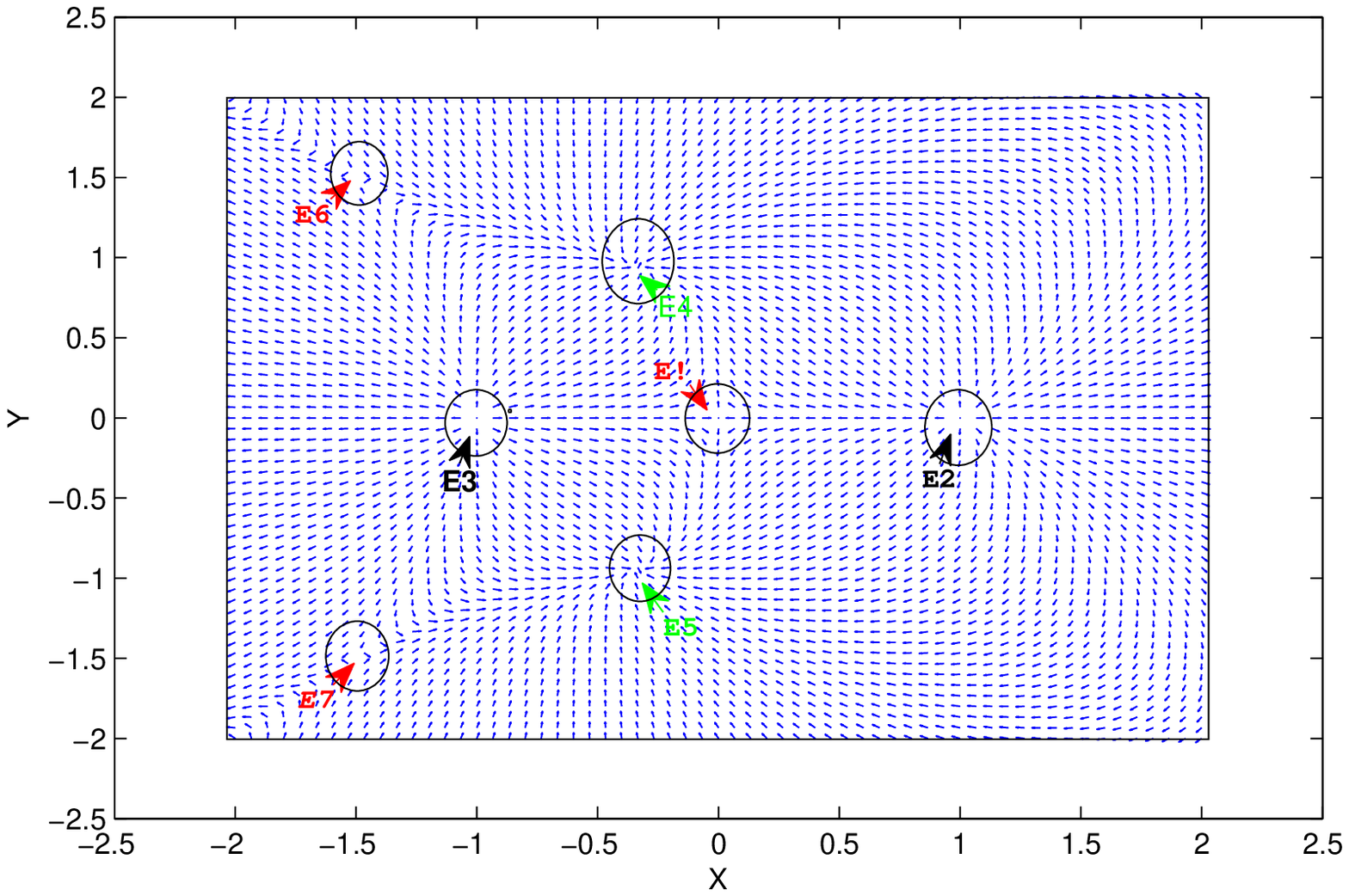}
  \caption{\label{}Direction field for the system given by (16) when $\alpha = 1 $, critical points (see table II) are indicated by different coloured arrows and nearby regions are circled. E1,E6,E7 are saddle point (red arrow), E4,E5 stable node (green arrow), E2,E3 unstable node (black arrow).}
 \end{minipage}
 \begin{minipage}{.45\textwidth}

 \includegraphics[width = .90\linewidth]{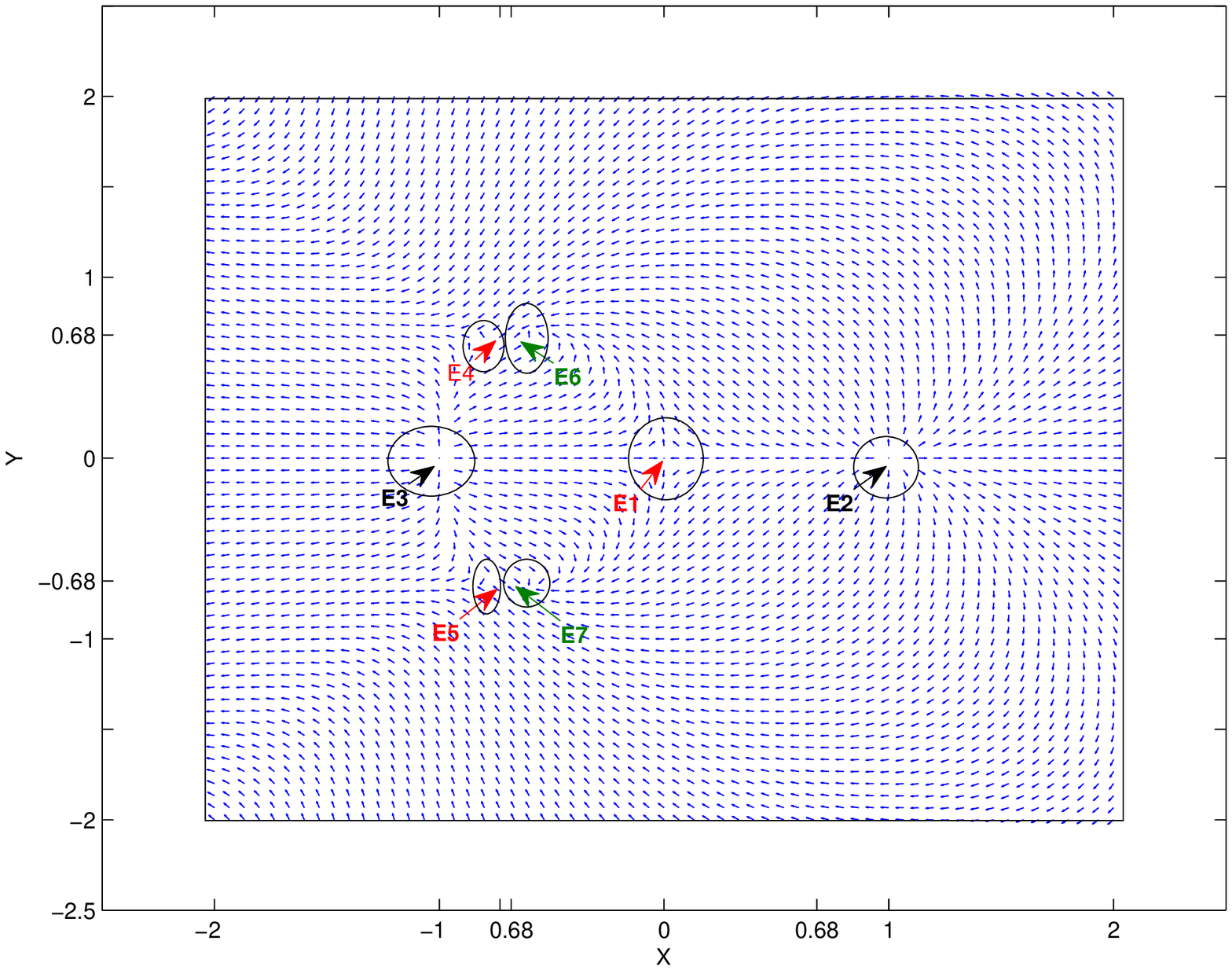}

 \caption{\label{}Direction field for the system given by(16) when $ \alpha = 2.2 $. Red arrow indicates saddle point (E1, E4, E5), black arrow unstable node (E2,E3), green arrow stable node (E6, E7).(see table II)}
 \end{minipage}
 \end{figure}
\subsection{ $ \alpha = \frac{1}{2x}$}
This choice of $ \alpha $ gives the scalar potential a simple form
\begin{equation}
 V = V_{0}a
\end{equation}
where 'a' is the scale factor of the FRW space-time metric. This choice of V will be realistic if 'a'is a function of $\phi$ . In particular, for exponential potential i.e $ V(\phi)= V_{0} e^{\mu\phi}$ the newly defined variable x turns out to be proportional to $\sqrt{\lambda(\phi)}$ , while for power law form of the potential function $ x \propto \phi\sqrt{\lambda(\phi)}$. In this case, the autonomous system (9) now simplifies to\\
$~~~~~~~ ~~~~~~~~~~~~~~~~~~~~~~~~~~~~~~~~~~~~~~~~~~~~~~~~~x'= \frac{3}{2} x (x^{2}- y^{2}-1)- \frac{ y^{2}}{2x} $
 \begin{equation}~~~~~~
 y'= \frac{y}{2}[4 + 3(x^{2}- y^{2})
\end{equation}
It is easy to see that (1, 0) and $(-1,0)$ are the only equilibrium points of the above mentioned autonomous system and both are unstable node in nature. Here both the equilibrium points correspond to massless scalar field or equivalently stiff fluid model.
\begin{figure}
 \includegraphics[width = .90\linewidth]{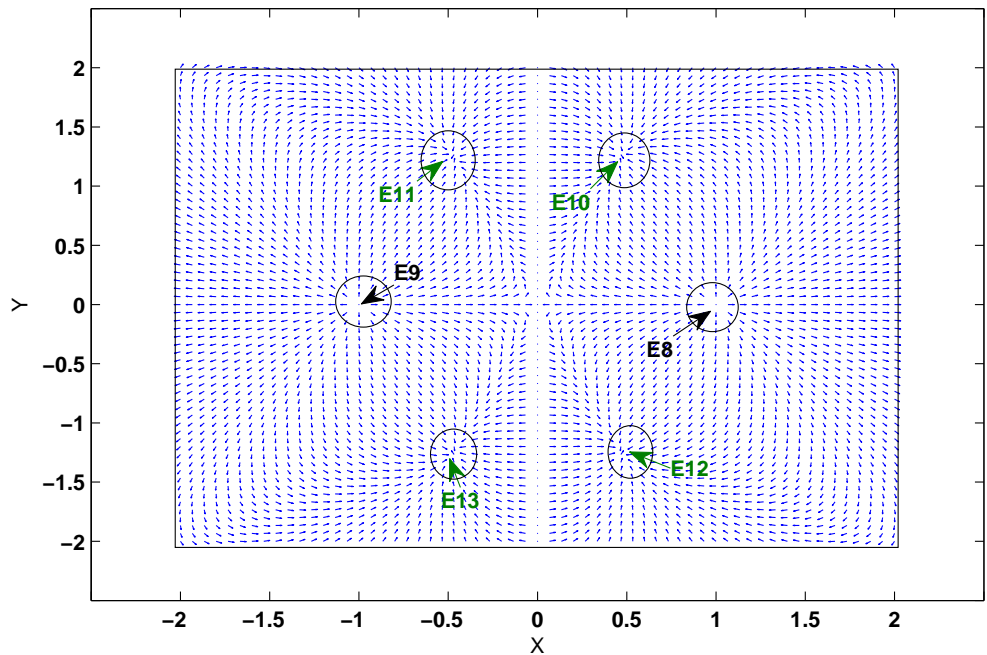}
\caption{\label{} Direction field for the system given by (9) when $ \alpha = -\frac{1}{2x}$. Critical points (see table III) are indicated by arrows.  E8, E9 unstable node (black arrow), E10, E11, E12, E13 stable node (green arrow). x = 0 is non allowable region.}
 \end{figure}
\subsection{  $ \alpha = - \frac{1}{2x}$ }
This choice of $ \alpha $ also corresponds to  a simple form of the scalar potential as
\begin{equation}
 V = \frac{V_{0}}{a}
\end{equation}

The autonomous system (16) now has six critical points which are given in table III. From the table, it is found that all the equilibrium points are purely dominated by dark energy scalar field. For the first two equilibrium points (which are unstable node) the scalar field represents ultra relativistic stiff fluid while remaining four equilibrium points are stable node and they represent accelerating universe in quintessence era and the scalar field behaves as a perfect fluid. Critical points and direction field near those points are shown in Fig 3.

\begin{table}
\caption{\emph{Equilibrium points, their nature and values of physical parameters  for $ \alpha = -\frac{1}{2x}$}}
 \begin{tabular}{|c|c|c|c|c|c|c|}
 \hline
 Equilibrium point & x & y & $  \Omega_{m} $ & $ \Omega_{\phi}$ & $ \nu_{\phi}$ & Nature\\\hline
 $ E_{8}$ & 1 & 0 & 0 & 1 & 2&  unstable node  \\\hline
 $ E_{9}$ & -1 & 0 & 0 & 1 & 2 & unstable node\\\hline
 $ E_{10}$ & $\frac{1}{\sqrt{6}}$ & $\sqrt{\frac{5}{6}}$ &0 & 1 & $ \frac{1}{3}$  & stable node \\\hline
 $E_{11}$ &  $-\frac{1}{\sqrt{6}}$ & $\sqrt{\frac{5}{6}}$ &0 & 1 & $ \frac{1}{3}$  & stable node\\\hline
 $E_{12}$ & $\frac{1}{\sqrt{6}}$ & $-\sqrt{\frac{5}{6}} $&0 & 1 & $  \frac{1}{3} $ & stable node\\\hline
 $E_{13}$ &  $-\frac{1}{\sqrt{6}}$ & $ -\sqrt{\frac{5}{6}}$ &0 & 1 & $\frac{1}{3}$ & stable node\\\hline

 \end{tabular}
 \end{table}

 \section{ Stability Criteria and equilibrium Points}
 For the present 2D autonomous system the local stability criteria of an equilibrium point is characterized by the eigen values of the perturbation matrix. In fact, the table IV shows the explicit criteria for the linear stability of the equilibrium points. We shall now investigate the classical as well as quantum stability of the model.

 %
\begin{table}
\caption{\emph{Eigen values of the linearized perturbation matrix and local stability}}
 \begin{tabular}{|p{2 in}|p{1.5 in}|p{1.5 in}|p{1.5 in}|}
 \hline
 \begin{center} Eigen values\end{center} &\begin{center} Trace of the linearized matrix\end{center} & \begin{center}Determinant of the perturbed matrix\end{center} & \begin{center}Nature of the critical point\end{center}\\\hline
  Eigen values with zero real part& \begin{center}0 \end{center}  & \begin{center}+ve \end{center}& \begin{center} Non-hyperbolic\end{center}\\\hline
  Eigen values with positive real part & \begin{center}+ve \end{center}& \begin{center} +ve \end{center}& \begin{center}Hyperbolic :           source  (unstable)\end{center}\\\hline
  Eigen values with negative real part &\begin{center} -ve\end{center} &  \begin{center}+ve \end{center}& \begin{center} Hyperbolic :             sink (stable)\end{center} \\\hline
  Eigen values with different signs &  \begin{center} indefinite\end{center} & \begin{center} -ve \end{center}& \begin{center} Hyperbolic : saddle\end{center}\\\hline
 \end{tabular}
 \end{table}

 In cosmological perturbations $ C_{s}^2$ appears as a coefficient of the term $~\frac{k^{2}}{a^{2}}$ ( k is the comoving momentum and 'a' is the usual scale factor) and classical fluctuations may be considered to be stable  when $ C_{s}^2$ is positive. For the quantum instabilities at UV scale we decompose the scalar field into a homogeneous part ($ \phi_{0}$) and a fluctuation as \\
 $ ~~~~~~~~~~~~~~~~~~~~~~~~~~~~~\phi (t,x) = \phi_{0}(t) + \delta \phi(t,x) $.\\
  Then  by expanding the pressure $ p(X,\phi$), upto second order in $\delta\phi$ , the Hamiltonian for the fluctuations  takes the form [24]\\
$ ~~~~~~~~~~~~~~~~~~~H = (p_{X} + 2Xp_{XX})\frac{(\delta\dot{\phi})^{2}}{2} + p_{X}\frac{(\nabla \delta\phi)^{2}}{2} - p_{\phi\phi}\frac{(\delta\phi)^{2}}{2}$\\
where suffix stands for differentiation with respect to the corresponding variable [25].\\
The above hamiltonian will be positive definite provided
\begin{equation}
   p_{X} + 2Xp_{XX} \geq 0  ,~~~~  p _{X} \geq 0 , ~~~~   - p_{\phi\phi} \geq 0
\end{equation}
and quantum stability is related to to the first two of the above conditions.

For the present problem,
 $ ~~~~C_{s}^2 = 1 + \alpha\frac{2y^2}{3x} ~~~$  and so for classical stability $ 3x^{2}+ 2 \alpha xy^{2}\geq 0 $ \\
for quantum stability \\ ~~~~~~~~~~~~~~~~~$3x^2 + 2 \alpha xy^2 \leq 0 $ ~~and~~ $~~~~~~ 18x^2 + 2 \alpha^{2} y^2 (y^2-2x^2) +3 \alpha xy^2 ( 3 + y^2 - x^2)\geq 0$, \\
We have shown both classical and quantum stability criteria of the model for the choices of $ \alpha $ in the table V.
\begin{table}
\caption{\emph{Stability Criteria of the Model}}
 \begin{tabular}{|p{1 in}|p{1 in}|p{1.2 in}|p{3in}|}
 \hline
  \begin{center} Choice of $\alpha$\end{center} & \begin{center}$C_{s}^{2}$ \end{center}& \begin{center}Classical stability \end{center} & \begin{center}Quantum stability \end{center} \\\hline
 \begin{center} Case -I \end{center} \begin{center} $\alpha $ constant \end{center} &  \begin{center} $ 1 + 2\alpha\frac{y^{2}}{3x}$ \end{center} &  \begin{center}$ 3x^{2}+ 2 \alpha xy^{2}\geq 0 $ \end{center} &  \begin{center} $3x^{2}+ 2 \alpha xy^{2}\leq 0$    and \end{center} \begin{center}$18x^2 + 2 \alpha^{2} y^2 (y^2-2x^2) +3 \alpha xy^2 ( 3 + y^2 - x^2)\geq 0$ \end{center} \\\hline
  \begin{center}Case-III \end{center}\begin{center}$ \alpha = -\frac{1}{2x}$ \end{center}&  \begin{center} $1 - \frac{y^{2}}{3x^{2}}$ \end{center} &  \begin{center}$ y^{2}\leq 3x^{2}$  \end{center}&  \begin{center}$ y^{2}\geq 3x^{2} $ and  \end{center} \begin{center} $ (y^{2} - 6x^{2})^{2}+ x^{2}y^{2}(1 + 3x^{2}-3y^{2})\geq 0 $ \end{center}\\\hline
 \end{tabular}
 \end{table}

We shall now examine whether criteria for model stability is obeyed at equilibrium points when x and y take the corresponding values of equilibrium points. From the table II, we see that we can not infer about the stability of the model at the critical point $ E_{1}$. The equilibrium points $ E_{2}$ and $ E_{3}$ are not locally stable and from the above model stability analysis, they correspond to classical stability but not quantum. The equilibrium points $ E_{4}$ and $ E_{5}$ correspond to classical stability of the model if $\alpha ^2 >\frac{9}{2}$ and as before they are not quantum stable. Finally, the equilibrium points $ E_{6}$ and $ E_{7}$ are classical (limiting) but not quantum stable. Similarly, from table III, we see that four stable equilibrium points $ E_{10}$, $ E_{11}$, $ E_{12}$ and $ E_{13}$ are both classical and quantum unstable. $ E_{8}$ and $ E_{9 }$ are not model stable.
\section{ Equilibrium points and cosmological implications}
From the above phase space analysis of the phantom scalar field model, we shall now discuss about the cosmological behaviour of the model at the equilibrium points. From table II, corresponding to the equilibrium point  $ E_{1}$, the solution corresponds to the standard Milne model and is not of much interest due to instability of the critical point. Both the critical points $ E_{2}$ and $ E_{3 }$ correspond to a massless scalar field or equivalently stiff perfect fluid FRW model and they are only classical stable. The equilibrium points $ E_{4}$ and $ E_{5}$ purely describe scalar field solutions in FRW model and are classically stable for the restriction $ \alpha^2 > \frac{9}{2}$ for which the critical points are unstable. $ E_{4 }$ represents late time attractor for $ \alpha^{2} < 3 $ and then critical point is stable but classical stability is  denied.  The remaining two critical points $ E_{6}$ and $ E_{7}$ are limiting stable only classically  and they represent  non accelerating matter scaling solution provided $\alpha ^2 >\frac{9}{2}$ for which  it is  locally stable (node) .

In table III, the critical points , $ E_{8}$ and $ E_{9}$ describe massless scalar field in FRW model and they are not of much interest as they are unstable node in nature. The remaining four critical points $ E_{10}$, $ E_{11}$, $ E_{12}$ and $ E_{13}$ are stable (node) from local linear analysis and at these points the potential energy dominates over the kinetic part and the scalar field behaves as exotic fluid. Thus, these critical points correspond to  cosmological solutions which describe the recently observed late time acceleration of the universe. It should be noted that although the critical points ($ E_{4}$, $ E_{5}$) or ($ E_{6}$, $ E_{7}$) have identical behaviour but from cosmological viewpoint they are not identical. As $ y > 0 $ for $ E_{4}$ and $ E_{6}$, so they correspond to expanding universe while $ E_{5}$ and $ E_{7}$ represent contracting universe. Similar is the situation for ($ E_{10}$, $ E_{12}$) and ($ E_{11}$, $ E_{13}$) in table III.

\section{ short discussion and conclusion}

The present work deals with a compact phase space analysis of a class of phantom scalar field cosmology, with the motivation of obtaining late time accelerating solutions compatible with observations. At first the evolution equations are reduced to a 3D autonomous system for exponential potential and exponential or power law coupling parameter . In both the cases the critical points are identical and are non-hyperbolic in nature . So linear stability analysis fails, only the dimension of the stable manifold is determined by central manifold theorem. Then we have studied the evolution equations  as 2D autonomous system for three choices of the phantom scalar field potential. We have obtained the critical points and presented the conditions for their existence and stability in tabular form for two choices of the scalar field potentials  while for the third choice of the potential we have only two critical points which do not correspond to realistic cosmological scenario and hence we have omitted them. The value of the physical parameters $\Omega_{\phi}$, $\nu _{\phi}$  and $\Omega _{m}$ are also presented in tables.\\
Further, we have investigated the classical as well as quantum stability of the model. Note that these two types of stability are not interrelated because the stability of a critical point is related to the perturbations $\delta x $ and $ \delta y $ ( and for the present 2D model the restriction is $ Tr  M < 0$ and $ det M > 0 $). On the other hand,the classical stability of the model is connected to the perturbations $\delta p$ (and depends on the condition $  C_{s}^2  \geq 0 $ ) while the quantum stability is related to the perturbations $\delta \phi$ and the conditions take the form of inequalities (21). Thus the critical points can be classified into three categories  namely\\
  (a) unstable points at which the model is stable,\\
  (b) stable points at which model is unstable and \\
  (c) stable points with stable ( both classical and quantum ) model. \\
  Obviously, the case (c) is interesting from cosmological point of view. From table II and III, we see that above (c) possibility is not valid for any critical point, although the critical points $ E_{4}$, $ E_{6} $ (in table II) and  two critical points in table III, namely $ E_{10}, E_{11}$   are interesting in the present context.\\
    Matter scaling solutions are not relevant attractors at late times. Thus we conclude that though there are late time attractors with phantom scalar field but they are  quantum mechanically unstable and is in agreement with the work of Ref [26].

\begin{acknowledgments}
The authors are thankful to IUCAA, Pune for warm hospitality and facilities at the library as major part of the work is done during a visit to IUCAA. The authors are also thankful to UGC-DRS programme, Department of Mathematics, Jadavpur University.
\end{acknowledgments}
\section{References}

\begin{thebibliography}{}
\bibitem{RefA}
S. J. Perlmutter et al., Astrophys. J. 517, 565 (1999); A. G. Riess et al., Astron. J. 116, 1009 (1998); C. L. Bennett et al., Astrophys. J. Suppl. Ser. 148, 1 (2003); D. J. Eisenstein et al., Astrophys. J. 633, 560 (2005); D. N. Spergel et al.(WMAP Collaboration), Astrophys. J. Suppl. Ser. 170, 377 (2007); A.G.Riess et al., Astrophys. J. 659, 98 (2007).
\bibitem{RefB}
E. Komatsu et al (WMAP Collaboration), Astrophys. J. Suppl. Ser 180, 330 (2009)
\bibitem{RefC}
S. M. Carroll, AIP Conf. Proc. 743, 16 (2004); E. J. Copeland, M. Sami and S. Tsujikawa , Int. J. Mod. Phys. D 15, 1753 (2006); V. Sahni and A. Starobinsky, Int. J. Mod. Phys. D 15, 2105 (2006)
\bibitem{RefD}
S. Weinberg, Rev. Mod. Phys. 61, 1 (1989)
\bibitem{RefE}
S. M. Carroll, W. H. Press and E. L. Turner, Ann. Rev. Astron. Astrophys. 30, 499 (1992)
\bibitem{RefF}
V. Sahni and A. Starobinsky, Int. J. Mod. Phys. D 9, 377 (2000)
\bibitem{RefI}
P. J. E. Peebles and B. Ratra, Rev. Mod. Phys. 75, 559 (2003)
\bibitem{RefJ}
T. Padmanabhan, Phys. Rep. 380, 235 (2003); Curr. Sci. 88, 1057 (2005)
\bibitem{RefK}
S. M. Carroll, Liv. Rev. Lett. 4, 1 (2001)
\bibitem{RefL}
L. Amendola and S. Tsujikawa, Dark Energy: Theory and Observations (Cambridge Univ. Press, Cambridge, England, 2010)
\bibitem{RefM}
R. R. Caldwell, Phys. Lett. B 545, 23 (2002)
\bibitem{RefN}
R. R. Caldwell, M.Kamionkowski and N. N. Weinberg, Phys. Rev. Lett. 91, 071301 (2003)
\bibitem{RefO}
R. J. Scherrer, Phys. Rev. D 71, 063519 (2005)
\bibitem{RefP}
J. Kujat, R. J. Scherrer and A. A. Sen, Phys. Rev. D 74, 083501 (2006)
\bibitem{RefB}
J. D. Barrow and C. G. Tsagas, Class. Quant. Grav. 26, 195003 (2009)
\bibitem{RefB}
J. Yoo and Y. Watanabe, Int. J. Mod. Phys. D 21, 1230002 (2012)
\bibitem{RefB}
L. P. Chimento, M. I. Forte, R. Lazcoz and M. G. Richarte, Phys. Rev. D 79, 043502 (2009)
\bibitem{RefB}
M. Hirsch and S. Smale, " Differential Equations, Dynamical Systems and Linear Algebra " (1974)(2nd edn, New york: Academic)
\bibitem{RefB}
G. Leon and E. N. Saridakis, J. Cosmol. Astropart. Phys. 11, 006 (2009); Phys. Lett. B 693, 1 (2010)
\bibitem{RefB}
X-m. Chen, Y. Gong and E. N. Saridakis, J. Cosmol. Astropart. Phys 28, 065012 (2011)
\bibitem{RefB}
R-J. Yong and X-T. Gao, Class. Quantum. Grav 04, 001 (2009)
\bibitem{RefB}
D. K. Arrowsmith and C. M. Place, " An Introduction  to Dynamical Syatems" (1990)( Cambridge Univ. Press, Cambridge)
\bibitem{RefB}
S. Wiggins, "Introduction to  Applied Nonlinear Dynamical Systems and Chaos" (2003)( 2nd Edn. Berlin; Springer)
\bibitem{RefB}
F. Piazza and S. Tsujikawa, JCAP 0407, 004 (2004)
\bibitem{RefB}
Usually, the fluctuations of a time-varying scalar obey Lorentz-violating dispersion relations. However, in cosmological context, Lorentz invariance is always violated considering preferred  ( CMB-, comoving observers-) frame.
\bibitem{RefB}
J. M. Cline, S. Jeon and G. D. Moore, Phys. Rev. D 70, 043543 (2004)
\end{thebibliography}

\end{document}